\documentclass[article]{IEEEtran}
\usepackage[utf8]{inputenc}
\usepackage{amsmath}
\usepackage{amssymb}
\usepackage{graphicx}
\usepackage{circuitikz}
\usepackage{tikz}
\usetikzlibrary{hobby}
\usepackage{cite}
\usepackage{xcolor}

\usepackage[greek,english]{babel}
\newcommand{\gpi}{\textrm{\greektext p}}
\renewcommand{\pi}{\gpi}

\renewcommand{\v}[1]{\boldsymbol{#1}}

\title{Radiation Efficiency Limits in Direct Antenna Modulation Transmitters}
\author{Kurt Schab \IEEEmembership{Member, IEEE}
	\thanks{Manuscript received  \today; revised \today.}
	\thanks{K.~Schab is with the Department of Electrical Engineering, Santa Clara University, Santa Clara, CA, USA (e-mail: kschab@scu.edu).}}
\begin{document}
	\maketitle
	
	\begin{abstract}
		Relative bounds on radiation efficiency are established for time-modulated antenna systems where radiation is generated by broadband conduction currents impressed using idealized non-radiating time-varying subsystems, such as those found in direct antenna modulation (DAM) transmitters.  Analytical and numerical examples demonstrate that the condition of quasi-resonance, common in nearly all practical direct antenna modulation transmitters, imposes severe restrictions on the otherwise unbounded gains in effective efficiency theoretically achievable by this class of time-modulated transmitters.
	\end{abstract}

	
	\section{Introduction}
	% DAM can beat Conv. ESA
	\label{sec:intro}
	\IEEEPARstart{T}{he} use of time modulation in an electrically small antenna or its matching network affords the possibility to exceed the strict physical bounds on its linear time invariant (LTI) performance.  Notably, direct antenna modulation (DAM) techniques based on the energy-synchronous modulation of time-varying components have been proposed, simulated, and measured as viable strategies for exceeding the conventional bandwidth-efficiency product limitations of small antennas, e.g., the Chu limit~\cite{Chu1948}.  Energy-synchronous DAM methods in the literature are typically specific to a signal type and antenna topology, with examples including on-off-keying for dipoles \cite{Galejs1963,Xu2006,schab2019pulse}, phase-shift-keying for dipoles \cite{schab2020phase}, and frequency shift keying on small loops \cite{Salehi2013,santos2019enabling}.  A theme common across these methods is the use of a time-varying matching network (varying on time scales much less than the carrier period) to impress broadband radiating currents onto the LTI portion of the antenna (e.g., metallic areas supporting conduction currents), thereby bypassing the filtering effects of the LTI radiator's impedance bandwidth. Each of the above techniques relies on tuning the LTI radiator to a quasi-resonant state, though other loop-based methods not requiring quasi-resonance have been reported~\cite{Merenda2006,manteghi2019}.
	
	% are these gains unbounded?  probably not since part of the system is still lti
	With the efficacy of DAM demonstrated by a variety of means, it now becomes necessary to determine whether or not there exist quantitative limits in performance gains achievable by adopting such techniques over traditional transmitters.  Here% we propose the relative efficiencies of DAM and conventional antennas transmitting an identical signal as a metric quantifying this performance advantage.  Restated, 
	~we formulate the question: for a fixed electrical size, how much more efficient could a DAM system be compared to an optimal resistively broadbanded conventional transmitter achieving the same effective bandwidth\footnote{Due to their dynamic nature, DAM transmitters do not have a classically defined impedance bandwidth.  An effective bandwidth may, however, be defined via a proxy measure such as distortion\cite{Lamensdorf1994,schab2019distortion}.}?  Defined in this way, it may appear that relative DAM efficiency increases indefinitely with decreasing transmitter size due to trends in radiation Q-factor.  However, upon closer inspection it appears that competing trends in efficiency may limit benefits in DAM efficiency gains in the extreme electrically small limit.  
	
	% goal and outline
	The goal of this letter is to calculate bounds on this relative efficiency gain by adapting tools developed for the analysis of classical LTI systems.  Namely, we examine low-frequency trends in the efficiency and Q-factor of small antennas to assess the relative benefit of idealized DAM systems over a broad range of scenarios (e.g., electrical size, tuning assumptions, conductivity models).  First, in Sec.~\ref{sec:models}, a DAM transmitter is decomposed into non-LTI and LTI components and models are proposed for ideal conventional, ideal non-resonant DAM, and ideal quasi-resonant DAM systems.  Then, in Sec.~\ref{sec:examples}, we apply these models to three examples of varying levels of abstraction and generality.  In all cases, we observe that the condition of quasi-resonance, much like that of self-resonance in classical antennas \cite{Smith_1977_TAP,Pfeiffer_FundamentalEfficiencyLimtisForESA,jelinek2018radiation}, imposes severe efficiency limitations and greatly reduces the potential gains afforded by DAM.%: a spherical shell for which optimal performance parameters are known analytically, a substructure (embedded antenna) problem which is analyzed numerically, and a driven wire dipole antenna common to several forms of DAM.
	
	\section{Modeling Ideal LTI and Direct Antenna Modulation Transmitters}
	\label{sec:models}
\begin{figure}
		\centering    
		\begin{circuitikz}[scale=0.8,transform shape]
			\draw (-2,1.25) to[short,-o](-1,1.25) to[short,o-] (0,1.25);
			\draw (-2,0.25) to[short,-o](-1,0.25) to[short,o-] (0,0.25);
			\path[draw,use Hobby shortcut,closed=true,scale=0.5,fill=yellow!20,thick]
			(0,0) .. (.5,1) .. (1.5,3) .. (.3,4) .. (-1,2) .. (-1,.5);
			\node at (0,1){$\boldsymbol{J}$};
			\draw (-6,1.25) to[short,-o](-4.5,1.25) to[short,o-] (-3,1.25);
			\draw (-6,0.25) to[short,-o](-4.5,0.25) to[short,o-] (-3,0.25);
			\draw[fill=white,thick] (-4,0) rectangle (-1.5,1.5) node[pos=.5,align=center] {LTI\\Matching};
			\draw[fill=white, thick] (-6.5,0) rectangle (-5,1.5) node[pos=.5,align=center] {Source};
		\end{circuitikz}\\
		\vspace{0.2in}
		
		\begin{circuitikz}[scale=0.8,transform shape]
			\draw (-2,1.25) to[short,-o](-1,1.25) to[short,o-] (0,1.25);
			\draw (-2,0.25) to[short,-o](-1,0.25) to[short,o-] (0,0.25);
			\draw (0,1.25) to[short,-o](1,1.25) to[short,o-] (1.5,1.25);
			\draw (0,0.25) to[short,-o](1,0.25) to[short,o-] (1.5,0.25);
			
			\draw[fill=white,thick] (1.5,0) rectangle (3.5,1.5) node[pos=.5,align=center] {non-LTI\\Loads};
			\path[draw,use Hobby shortcut,closed=true,scale=0.5,fill=yellow!20,thick]
			(0,0) .. (.5,1) .. (1.5,3) .. (.3,4) .. (-1,2) .. (-1,.5);
			\node at (0,1){$\boldsymbol{J}$};
			\draw (-6,1.25) to[short,-o](-4.5,1.25) to[short,o-] (-3,1.25);
			\draw (-6,0.25) to[short,-o](-4.5,0.25) to[short,o-] (-3,0.25);
			\draw[fill=white,thick] (-4,0) rectangle (-1.5,1.5) node[pos=.5,align=center] {non-LTI\\Matching};
			\draw[fill=white, thick] (-6.5,0) rectangle (-5,1.5) node[pos=.5,align=center] {Source};
		\end{circuitikz}
		
		\caption{Conventional (top) and time-modulated (bottom) transmitters.  In both setups, the impressed current distribution $\v{J}$ is the only source of radiation.}
		\label{fig:schem}
	\end{figure}
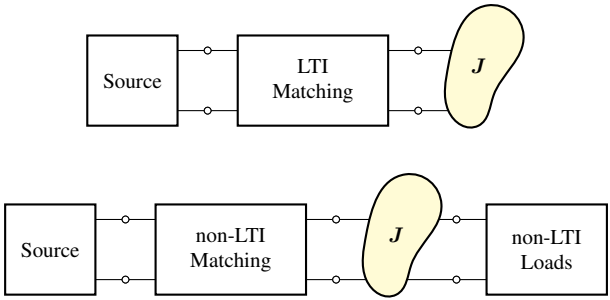

	Throughout this paper we assume that conventional and DAM systems are represented by the block diagrams in Fig.~\ref{fig:schem}.  In both systems, radiation is produced only by conduction currents induced on the object $\varOmega$.  We do not consider non-conduction-based radiation, e.g., acoustically driven antennas \cite{hassanien2020theoretical}.  Whether driven by LTI or non-LTI systems, the physical bounds governing the maximum achievable radiation efficiency by these currents remain the same.  However, when driven by non-LTI systems, the impact of high Q-factor may not impair the realized bandwidth of a DAM transmitter.
	
\begin{figure}
		\centering
		\includegraphics[width=3.25in]{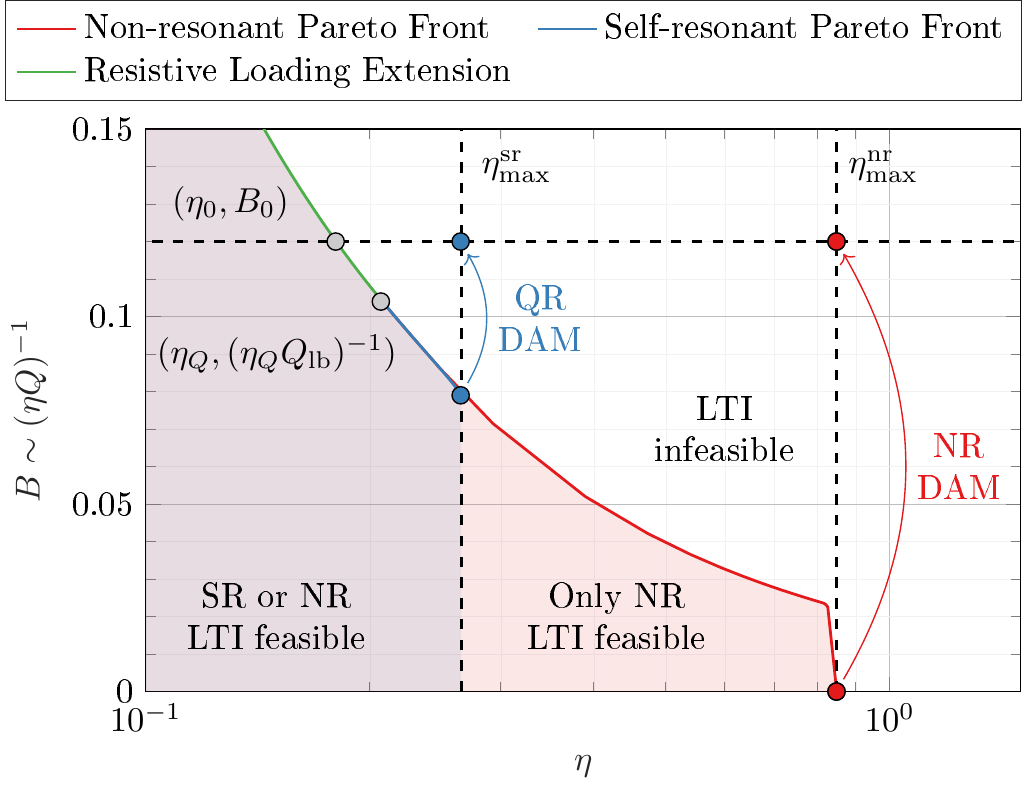}
		\caption{Graphical representation of calculating the effective efficiency gains of a non-resonant (NR) and quasi-resonant (QR) DAM transmitter using Pareto-optimal current analysis.  Specifications for the example generating these data are taken from \cite[\S VIII-A]{gustafsson2019trade}.}
		\label{fig:pareto}
	\end{figure}

	The tradeoff between optimal efficiency and bandwidth for current distributions confined to the design region $\varOmega$\footnote{By proxy, this includes all possible antennas confined to the region $\varOmega$~\cite{gustafsson2016antenna}.} may be rigorously calculated via the multiobjective optimization methods described in \cite{gustafsson2019trade}.  Examples of these tradeoffs, in the form of Pareto fronts \cite{boyd2004convex}, are shown in Fig.~\ref{fig:pareto}. Two Pareto fronts are plotted with and without the requirement of self-resonance, with the latter being less restrictive.  Points in bandwidth-efficiency space above each Pareto front are infeasible by LTI transmitters.  The self-resonant and non-resonant Pareto fronts meet at the coordinate $(\eta_Q, (\eta_Q Q_\mathrm{lb})^{-1})$ corresponding to the minimum achievable radiation Q-factor.  From this point both Pareto fronts may be extended toward higher bandwidth via resistive loading, represented by the curve segment $B = (\eta,(\eta Q_\mathrm{lb})^{-1})$ with $\eta \leq \eta_Q$.  For a prescribed bandwidth $B_0$, the maximum efficiency $\eta_0$ achievable by an LTI antenna is readily obtained by finding the intersection of the line $B = B_0$ and the chosen Pareto front, as shown in Fig.~\ref{fig:pareto}.  Using this graphical representation of antenna bounds, we now describe models for ideal LTI transmitters, as well as two models for ideal DAM transmitters.
	
	\subsection{Ideal Conventional LTI Transmitter}
	We assume that a conventional antenna is matched to a real source impedance via single resonance tuning, see Fig.~\ref{fig:schem}.  This tuning may be carried out either via antenna shaping or through the use of external lumped elements.  In either case, we assume that the tuning network is constructed of the same materials as the antenna itself and that the tuning network is confined to the design region of the antenna.  Under these assumptions, we observe that the maximum efficiency possible for a conventional LTI antenna is given by the intersection of the $B = B_0$ line and the self-resonant Pareto front, i.e.,
	\begin{equation}
	\eta_\mathrm{Conv.} = \min\{\eta_0,\eta^\mathrm{sr}_\mathrm{max}\},
	\end{equation}
	where $\eta_0$ is the efficiency at the intersection and minimization explicitly enforces the right boundary of the self-resonant Pareto front in Fig.~\ref{fig:pareto}.
	
	\subsection{Ideal Non-resonant DAM Transmitter}
	Using non-LTI matching and loading networks, it may be possible to impress a broadband signal onto the maximally efficient current distribution feasible within the LTI design region $\varOmega$.  If this to be (at least hypothetically) true, then an ideal DAM transmitter with no constraint on resonance or quasi-resonance would the achieve arbitrary bandwidth $B_0$ with efficiency $\eta_\mathrm{DAM}^\mathrm{nr} = \eta^\mathrm{nr}_\mathrm{max}$.  The actual topology of the antenna and non-LTI system that would achieve this performance is unknown, but this proposed model serves as a first upper bound on DAM performance.
	
	\subsection{Ideal Quasi-resonant DAM Transmitter}
	The aforementioned non-resonant upper bound on radiation efficiency is known to be loose.  Additionally, many DAM transmitter architectures rely on resonant tuning to achieve quasi-resonance, e.g., \cite{Galejs1963}, as part of their broadbanding strategy.  Here we incorporate these features by assuming that an ideal quasi-resonant DAM transmitter may realize arbitrary bandwidth $B_0$ by impressing a broadband version of the maximum self-resonant efficiency current distribution on the design region $\varOmega$, i.e., $\eta_\mathrm{DAM}^\mathrm{qr} = \eta^\mathrm{sr}_\mathrm{max}$.  The example data in Fig.~\ref{fig:pareto} representatively demonstrate that the self-resonant efficiency bound is lower than the non-resonant bound \cite{Smith_1977_TAP,Pfeiffer_FundamentalEfficiencyLimtisForESA,jelinek2018radiation}.
	
	\section{Effective Efficiency Gains}
	\label{sec:examples}
	With models for the ideal efficiency of conventional, non-resonant DAM, and quasi-resonant DAM transmitters established, we proceed by studying the relative efficiency of each DAM transmitter with respect to the conventional case, i.e., $\eta_\mathrm{DAM}^{\mathrm{nr}} / \eta_\mathrm{Conv.}$ and  $\eta_\mathrm{DAM}^{\mathrm{qr}} / \eta_\mathrm{Conv.}$.  These quotients represent the highest possible relative gain in efficiency feasible by each class of DAM transmitter over a conventional LTI system. In the remainder of this section we study these ratios for three examples with unique characteristics.  %First, a spherical shell with analytically known Pareto fronts is studied.  Next, a more complex example is examined where the Pareto fronts must be calculated numerically.  Finally, the analysis is modified to study a driven wire dipole antenna common in a certain class of energy-synchronous DAM transmitters. 
	
	\subsection{Electrically Small Spherical Shell}
	\label{sec:ex-sphere}
	As a first example, we analytically study the optimal performance of a spherical shell carrying a TM$_{10}$ (dipole) modal current distribution.  Asymptotic forms for the Q-factor and dissipation factor\footnote{Asymptotic scaling rules for loss for this example are much clearer when expressed in terms of dissipation factor $\delta$, related to efficiency via \mbox{$\eta = (1+\delta)^{-1}$}, see~\cite{jelinek2018radiation,gustafsson2019trade}.} of this structure in the electrically small ($ka\rightarrow 0$) limit are \cite{Losenicky-2018-DissipationFactorsOfSphericalCurrentModes}
	\begin{subequations}
		\begin{equation}
		Q_\mathrm{rad}^\mathrm{nr} = \frac{3}{2(ka)^3},\quad Q_\mathrm{rad}^\mathrm{sr} = \frac{1}{(ka)^3},
		\end{equation}
		\begin{equation}
		\delta^\mathrm{nr} = \frac{R_\mathrm{s}}{Z_0}\frac{9}{4(ka)^2},\quad \delta^\mathrm{sr} = \frac{R_\mathrm{s}}{Z_0}\frac{3}{(ka)^4},
		\end{equation}
		\label{eq:sphere-exp}
	\end{subequations}
	where $R_\mathrm{s}$ and $Z_0$ are the shell's surface resistivity and the impedance of free space, respectively.  The non-resonant case is optimal in non-resonant dissipation factor, while the self-resonant case (where the TE$_{10}$ mode is used for tuning) represents both minimum Q-factor and minimum self-resonant dissipation factor.  It can be shown that the Pareto front between these parameters is linear in $\delta$-$Q_\mathrm{rad}$ space between these points \cite{gustafsson2019trade}.  
	
\begin{figure}
		\centering
		%\testfig
		\includegraphics[width=3.25in]{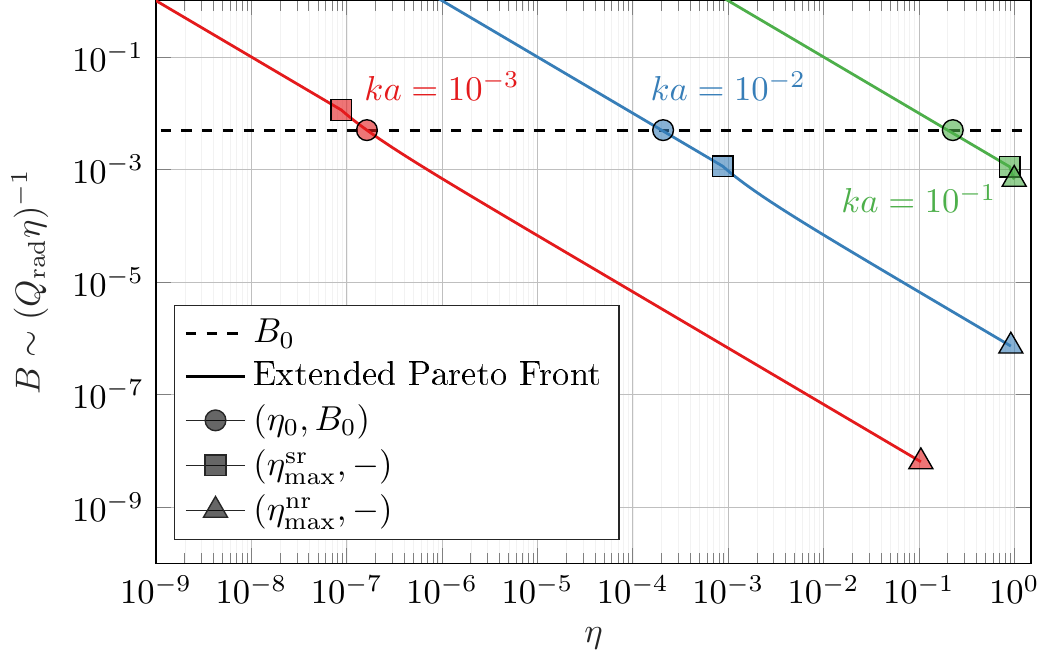}
		\caption{Analytic Pareto fronts for optimal spherical transmitters.  Square and triangular markers denote the self-resonant and non-resonant efficiency bounds.  Circles denote the intercept between each Pareto front and the objective bandwidth $B_0 = 0.005$.}
		\label{fig:sphere-pareto}
	\end{figure}

	Pareto fronts for shells of three electrical sizes are shown in Fig.~\ref{fig:sphere-pareto}.  For this example calculation, a non-dispersive surface resistivity $R_\mathrm{s} = 0.0014~\Omega/\square$ is used, corresponding to thick copper at $30~\mathrm{MHz}$~\cite{Balanis1982}, and an objective bandwidth is set at $B_0 = 0.005$.  The maximal efficiencies $\eta^\mathrm{nr}_\mathrm{max}$ and $\eta^\mathrm{sr}_\mathrm{max}$ are denoted by square and triangular markers on each Pareto front. The intercept between each Pareto front and the \mbox{$B=B_0$} line are also marked. These data show that for all three electrical sizes, the non-resonant DAM efficiency is much higher than the maximally efficient conventional transmitter.  This is not the case for DAM transmitters with the constraint of quasi-resonance, which have efficiency that reduces rapidly with decreasing electrical size, eventually passing to the left of the Pareto intercept point $(\eta_0,B_0)$.
	
\begin{figure}
		\centering
		%\testfig
		\includegraphics[width=3.25in]{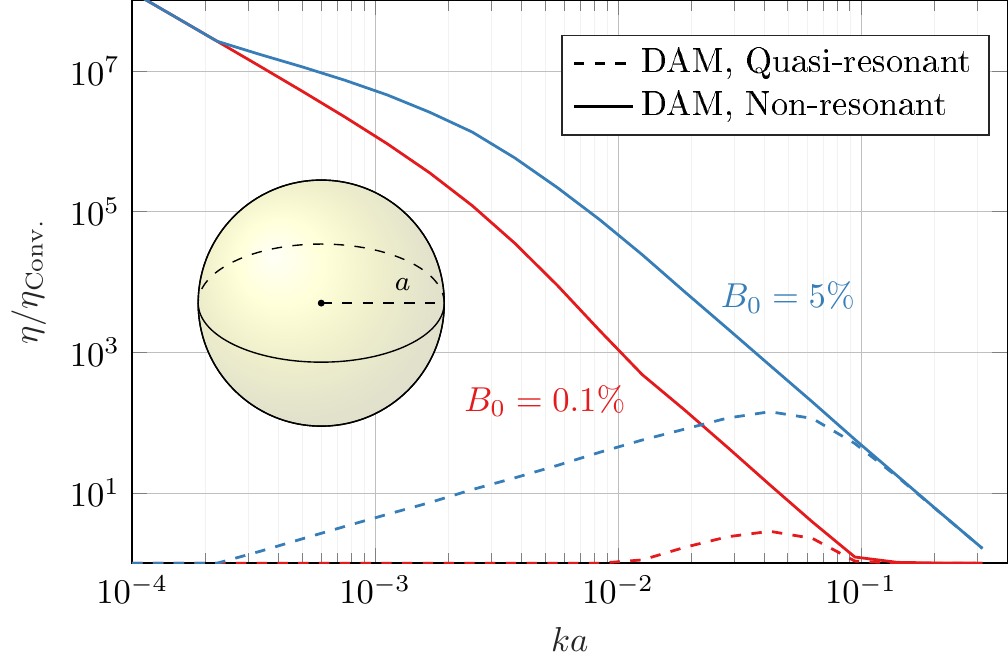}
		\caption{Effective efficiency gains for optimal spherical transmitters.}
		\label{fig:sphere-eff}
	\end{figure}

	The trends observed in Fig.~\ref{fig:sphere-pareto} indicate that the ``effective efficiency gain'' $\eta / \eta_\mathrm{Conv}$ in a DAM system depends highly on whether or not the condition of quasi-resonance is enforced.  At moderately small electrical sizes, gains in both forms of DAM are substantial.  For extremely small electrical sizes, the rapid increase in losses imposed by resonance overtakes the necessary efficiency sacrifice required to resistively broadband a conventional transmitter to the objective bandwidth $B_0$.  This is further demonstrated in Fig.~\ref{fig:sphere-eff}, where the effective efficiency gain $\eta/\eta_\mathrm{Conv.}$ is plotted as a function of electrical size $ka$ for two objective bandwidths.  There we clearly observe that gains for non-resonant DAM transmitters are unbounded in the electrically small limit, whereas quasi-resonant DAM transmitters provide an efficiency benefit only over a limited size range.  Naturally, the span of this range depends on the target bandwidth, with more extreme broadbanding leading to gains over broader ranges.
	
	\subsection{Optimal Substructure Embedded Antenna}
	\label{sec:ex-substructure}
\begin{figure}
		\centering
		%\testfig
		\includegraphics[width=3.25in]{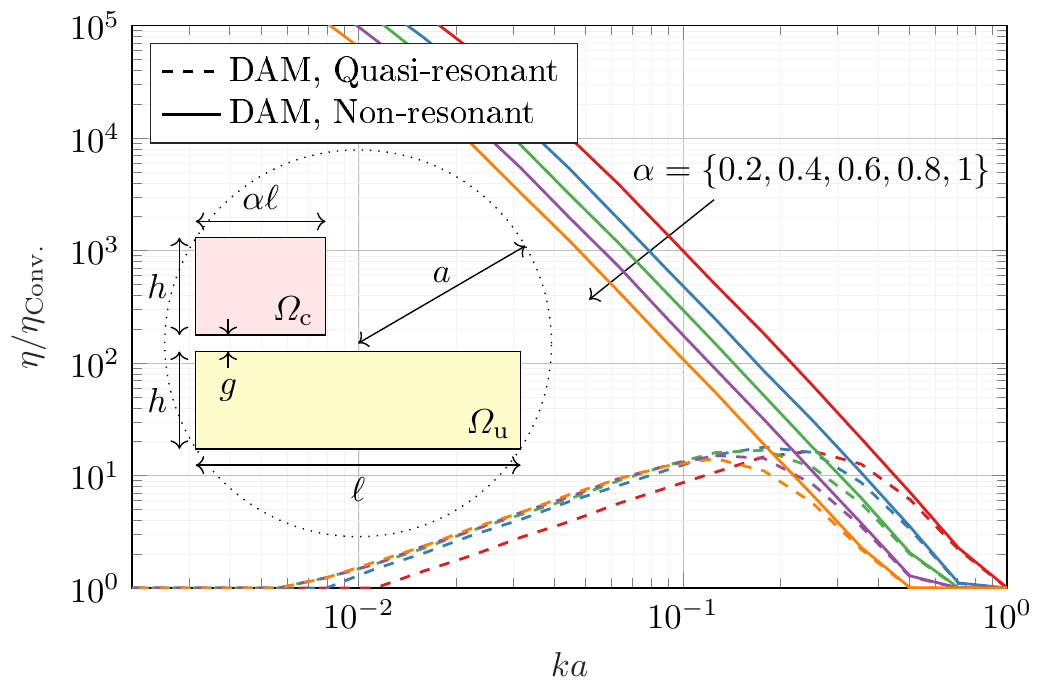}
		\caption{Effective efficiency of a substructure embedded system.  Optimal currents are calculated using the method of moments based techniques in \cite{gustafsson2019trade} using AToM \cite{}.  The controllable region $\varOmega_\mathrm{c}$ is scaled by the parameter $\alpha$ while the uncontrollable region $\varOmega_\mathrm{u}$ is kept fixed.  In this example, $h = 0.3\ell$ and $g = 0.05\ell$.}
		\label{fig:controllable}
	\end{figure}

	As a more complex example utilizing the generality of the Pareto front calculation methods in \cite{gustafsson2019trade}, an embedded antenna is considered where only a portion of the current support is considered controllable.  The controllable region $\varOmega_\mathrm{c}$ over which currents may be optimized is defined as a rectangular region of dimensions $h \times \alpha \ell$.  The uncontrollable scatterer $\varOmega_\mathrm{u}$ is located a distance $g$ below the controllable region and has dimensions $h \times \ell$, as shown in Fig.~\ref{fig:controllable}.  As in the previous example, a non-dispersive surface resistance  $R_\mathrm{s} = 0.0014~\Omega/\square$ is used.  

	Effective efficiency gains for several different values of the parameter $\alpha$ are shown in Fig.~\ref{fig:controllable}, where we observe the same trends as in the spherical transmitter.  Interestingly, the low-frequency trend in quasi-resonant DAM efficiency gain appears to be only weakly dependent on the geometry parameter $\alpha$.  This can be understood as a rather weak dependence of the maximally efficiency tuning current Q-factor $Q_L$ on this parameter, consistent with the competing trends in increasing electrical size and decreasing aspect ratio for standalone rectangular plates \cite{gustafsson2019trade}.

	\subsection{Driven Wire Dipole}
	As a final example, we step away from Pareto-optimal current analysis and calculate the relative efficiencies of a driven antenna.  Here we explicitly require conventional and DAM systems to radiate via the dipole's driven current distribution, though DAM systems may be designed to impress this current distribution over a broad bandwidth, bypassing the dipole's narrow impedance bandwidth (see \cite{Galejs1963}).  The impedance \mbox{$Z_\mathrm{a} = R_\mathrm{a} +\mathrm{j}X_\mathrm{a}$} and radiation efficiency $\eta_\mathrm{a}$ of a center-fed copper wire dipole with length $2~\mathrm{m}$ and wire radius $1.6~\mathrm{mm}$ are calculated using NEC \cite{nec} over the frequency range $0.2$ -- $70~\mathrm{MHz}$.  %Asymptotic forms for the dipole's impedance \mbox{$Z_\mathrm{a} = R_\mathrm{a} +\mathrm{j}X_\mathrm{a}$} and efficiency $\eta_\mathrm{a}$ are available in closed form~\cite{Balanis1982} and we discuss their behavior at the end of this section.
	
	For a conventional transmitter, the antenna may be tuned to resonance at each frequency via an inductance $L$ with Q-factor $Q_L$, giving rise to the total efficiency 
	\begin{equation}
	\eta =  \frac{\eta_\mathrm{a}R_\mathrm{a}}{R_\mathrm{a}-X_\mathrm{a}/Q_L}.
	\label{eq:dipole_eta_tuned}
	\end{equation}
	As in the previous example, the conventional efficiency is assumed to be tuned and resistively loaded as needed to obtain a target bandwidth $B_0$, i.e.,
	\begin{equation}
	\eta_\mathrm{Conv.} = \min \{(B_0 Q_Z)^{-1},\eta\},
	\end{equation}
	where $Q_Z$ is the impedance-based Q-factor of the standalone\footnote{Calculating $Q_Z$ requires implicit inclusion of a lossless tuning element in conjunction with measured or simulated impedance data.} antenna \cite{Yaghjian2005}.
		
	An ideal DAM transmitter which requires the same resonant tuning as in the conventional system (see \cite{Galejs1963, Xu2010,schab2019pulse}) will, at best, achieve the resonant efficiency $\eta_\mathrm{DAM}^\mathrm{qr} = \eta$, whereas one without the resonance constraint would achieve the standalone antenna efficiency $\eta_\mathrm{DAM}^\mathrm{nr} = \eta_\mathrm{a}$.  Note that the latter is equivalent to assuming lossless inductors are available, though rigorous comparison to the conventional case in that particular scenario requires an alteration to \eqref{eq:dipole_eta_tuned}, so we do not consider this interpretation further in the present paper.  Taking the ratio of conventional and DAM transmitter efficiencies, with some rearranging, yields
	\begin{subequations}
		\begin{equation}
		\eta_\mathrm{DAM}^\mathrm{qr} / \eta_\mathrm{Conv.} = \max \bigg \{\frac{Q_Z R_\mathrm{a} B_0}{R_\mathrm{a} - X_\mathrm{a}/Q_L},1\bigg \}
		\label{eq:dipole-sr}
		\end{equation}
		\begin{equation}
		\eta_\mathrm{DAM}^\mathrm{nr} / \eta_\mathrm{Conv.} = \max \bigg \{Q_Z B_0,\frac{R_\mathrm{a} - X_\mathrm{a}/Q_L}{R_\mathrm{a}}\bigg \}.
		\end{equation}
		\label{eq:dipole-eff}
	\end{subequations}
	
	\begin{figure}
		\centering
		%\testfig
		\includegraphics[width=3.25in]{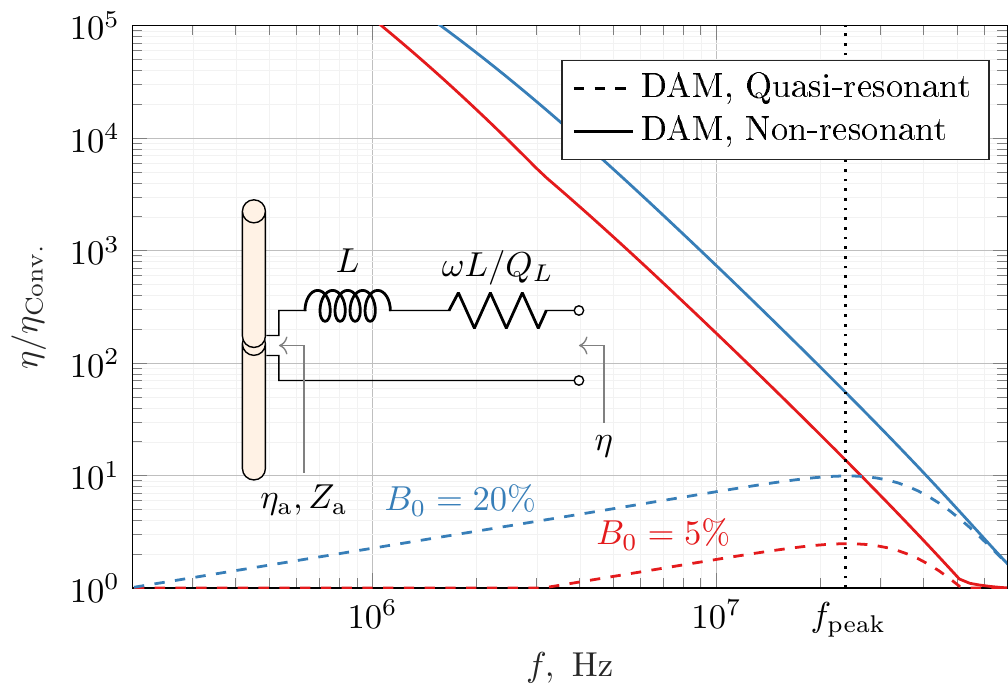}
		\caption{Effective efficiency gains of a wire dipole antenna.}
		\label{fig:dipole}
	\end{figure}
	
	Curves are shown in Fig.~\ref{fig:dipole} illustrating the efficiency ratios in \eqref{eq:dipole-eff} for two bandwidths $B_0$ using tuning elements defined by a realistic frequency dependent Q-factor for air-coil inductors \cite{coilcraft:square-air-coil,jelinek2018radiation}
	\begin{equation}
	Q_L = 150 \sqrt{\omega / (2\pi\cdot 30\cdot 10^{6})}.
	\label{eq:ql}
	\end{equation}
	We observe the same characteristic features as in the previous examples, where efficiency gains in the non-resonant system are potentially unbounded in the low frequency limit while gains in the resonant system are limited and present only over a finite bandwidth.  This example indicates that for realistic loss models, physical size, and frequency ranges, the potential benefits of a DAM transmitter are severely limited when some form of resonance (e.g., quasi-resonance) is required.
	
	In the electrically small limit, the dipole antenna's impedance behavior is accurately\footnote{Note that the skin depth model is used throughout, which is non-physical in the extreme electrically small limit.} approximated as
	\begin{equation}
	R_\mathrm{a} \approx R_{\mathrm{rad},0} \frac{\omega^2}{\omega_0^2} + R_{\Omega,0}\sqrt{\frac{\omega}{\omega_0}},\quad X_\mathrm{a} \approx \frac{1}{\mathrm{j}\omega C_0}
	\label{eq:dipole-zin}
	\end{equation}
	where $R_{\mathrm{rad},0}$, $R_{\Omega,0}$, and $C_0$ are the radiation resistance at frequency $\omega_0$, the ohmic loss resistance at frequency $\omega_0$, and static capacitance, respectively.  Additionally, the antenna's radiation Q-factor may be modeled with the asymptotic cubic scaling in electrical size, i.e., $Q_\mathrm{a} = Q_\mathrm{a,0}\omega_0^3/\omega^3$, and a general form of \eqref{eq:ql} where $Q_L = Q_{L,0}\sqrt{\omega/\omega_0}$ may be assumed for the inductor Q-factor.  Determining the frequency range over which the self-resonant DAM efficiency in \eqref{eq:dipole-sr} is greater than unity in this case is best carried out numerically, however the location of the peak value of self-resonant efficiency gain may be estimated as
	\begin{equation}
	\omega_\mathrm{peak} = \sqrt[\leftroot{-1}\uproot{2}\scriptstyle 7]{\frac{\omega_0^{5}}{\left(6R_{\mathrm{rad},0}C_0Q_{L,0}\right)^{2}}}
	\end{equation}
	by assuming that at this frequency radiation and inductor losses far outweigh ohmic losses.  In Fig.~\ref{fig:dipole} we observe that this approximation, which results in a peak location dependent only on low frequency asymptotic coefficients, is extremely accurate ($<1\%$ error).  A similar analysis may be carried out by any small dipole radiator characterized by \eqref{eq:dipole-zin}.  %Including on-antenna tuning elements effectively returns the problem to one of studying Pareto-optimal current distributions, as in Secs.~\ref{sec:ex-sphere} and \ref{sec:ex-substructure}.

	\section{Conclusions}
	In this paper, an effective efficiency ratio is used to quantify the advantage of DAM transmitters over their conventional counterparts.  By splitting the system into LTI and non-LTI components, many results from optimal LTI antenna theory are applied.  From all reported examples, we observe that the requirement of resonance (e.g., quasi-resonance) severely limits the potential benefits of a DAM transmitter.  This implies that quasi-resonant DAM systems, prevalent in the literature, may have limited applicability at extremely low frequencies.  The same conclusion holds for any loop based methods, regardless of resonance condition, as their radiation efficiency scales inherently as $(ka)^4$.  Similar concepts may be applied to the analysis of scattering from non-linear loads \cite{liu1976nonlinear,landt1983agard,palmer2019investigation}, as the underlying bifurcation of a system into linear and non-linear components is very similar to that used in the antenna problems studied here.
	
	\bibliographystyle{IEEEtran}
	\bibliography{main}

\begin{thebibliography}{10}
\providecommand{\url}[1]{#1}
\csname url@samestyle\endcsname
\providecommand{\newblock}{\relax}
\providecommand{\bibinfo}[2]{#2}
\providecommand{\BIBentrySTDinterwordspacing}{\spaceskip=0pt\relax}
\providecommand{\BIBentryALTinterwordstretchfactor}{4}
\providecommand{\BIBentryALTinterwordspacing}{\spaceskip=\fontdimen2\font plus
\BIBentryALTinterwordstretchfactor\fontdimen3\font minus
  \fontdimen4\font\relax}
\providecommand{\BIBforeignlanguage}[2]{{%
\expandafter\ifx\csname l@#1\endcsname\relax
\typeout{** WARNING: IEEEtran.bst: No hyphenation pattern has been}%
\typeout{** loaded for the language `#1'. Using the pattern for}%
\typeout{** the default language instead.}%
\else
\language=\csname l@#1\endcsname
\fi
#2}}
\providecommand{\BIBdecl}{\relax}
\BIBdecl

\bibitem{Chu1948}
L.~J. {Chu}, ``{Physical Limitations of Omni-Directional Antennas},''
  \emph{Journal of Applied Physics}, vol.~19, pp. 1163--1175, Dec. 1948.

\bibitem{Galejs1963}
J.~Galejs, ``Switching of reactive elements in high-{Q} antennas,'' \emph{IEEE
  Transactions on Communications Systems}, vol.~11, no.~2, pp. 254--255, June
  1963.

\bibitem{Xu2006}
X.~Xu, H.~C. Jing, and Y.~E. Wang, ``High speed pulse radiation from switched
  electrically small antennas,'' in \emph{Antennas and Propagation Society
  International Symposium 2006, IEEE}.\hskip 1em plus 0.5em minus 0.4em\relax
  IEEE, 2006, pp. 167--170.

\bibitem{schab2019pulse}
K.~Schab, D.~Huang, and J.~J. Adams, ``Pulse characteristics of a direct
  antenna modulation transmitter,'' \emph{IEEE Access}, 2019.

\bibitem{schab2020phase}
K.~{Schab}, D.~{Huang}, and J.~J. {Adams}, ``An energy-synchronous direct
  antenna modulation method for phase shift keying,'' \emph{IEEE Open Journal
  of Antennas and Propagation}, pp. 1--1, 2020.

\bibitem{Salehi2013}
M.~Salehi, M.~Manteghi, S.-Y. Suh, S.~Sajuyigbe, and H.~G. Skinner, ``A
  wideband frequency-shift keying modulation technique using transient state of
  a small antenna,'' \emph{Progress In Electromagnetics Research}, vol. 143,
  pp. 421--445, 2013.

\bibitem{santos2019enabling}
J.~P. Santos, F.~Fereidoony, and Y.~E. Wang, ``Enabling high efficiency
  bandwidth electrically small antennas through direct antenna modulation,'' in
  \emph{2019 IEEE International Symposium on Antennas and Propagation and
  USNC-URSI Radio Science Meeting}.\hskip 1em plus 0.5em minus 0.4em\relax
  IEEE, 2019, pp. 1539--1540.

\bibitem{Merenda2006}
J.~T. Merenda, ``Digital wideband small antenna systems,'' \emph{BAE Systems},
  2006.

\bibitem{manteghi2019}
M.~{Manteghi}, ``Fundamental limits, bandwidth, and information rate of
  electrically small antennas: Increasing the throughput of an antenna without
  violating the thermodynamic {Q-}factor,'' \emph{IEEE Antennas and Propagation
  Magazine}, pp. 1--1, 2019.

\bibitem{Lamensdorf1994}
D.~Lamensdorf and L.~Susman, ``Baseband-pulse-antenna techniques,'' \emph{IEEE
  Antennas and Propagation Magazine}, vol.~36, no.~1, pp. 20--30, Feb 1994.

\bibitem{schab2019distortion}
K.~{Schab}, A.~{Singh}, and N.~{Bohannon}, ``Distortion analysis for the
  assessment of {LTI} and non-{LTI} transmitters,'' \emph{IEEE Transactions on
  Antennas and Propagation}, pp. 1--1, 2020.

\bibitem{Smith_1977_TAP}
G.~S. Smith, ``Efficiency of electrically small antennas combined with matching
  networks,'' \emph{IEEE Trans. Antennas Propag.}, vol.~25, pp. 369--373, 1977.

\bibitem{Pfeiffer_FundamentalEfficiencyLimtisForESA}
C.~Pfeiffer, ``Fundamental efficiency limits for small metallic antennas,''
  \emph{IEEE Trans. Antennas Propag.}, vol.~65, pp. 1642--1650, 2017.

\bibitem{jelinek2018radiation}
L.~Jelinek, K.~Schab, and M.~Capek, ``Radiation efficiency cost of resonance
  tuning,'' \emph{IEEE Transactions on Antennas and Propagation}, 2018.

\bibitem{hassanien2020theoretical}
A.~E. Hassanien, M.~Breen, M.-H. Li, and S.~Gong, ``A theoretical study of
  acoustically driven antennas,'' \emph{Journal of Applied Physics}, vol. 127,
  no.~1, p. 014903, 2020.

\bibitem{gustafsson2016antenna}
M.~Gustafsson, D.~Tayli, C.~Ehrenborg, M.~Cismasu, and S.~Nordebo, ``Antenna
  current optimization using matlab and cvx,'' \emph{FERMAT}, vol.~15, no.~5,
  pp. 1--29, 2016.

\bibitem{gustafsson2019trade}
M.~Gustafsson, M.~Capek, and K.~Schab, ``Trade-off between antenna efficiency
  and {Q}-factor,'' \emph{IEEE Transactions on Antennas and Propagation}, 2019.

\bibitem{boyd2004convex}
S.~Boyd, S.~P. Boyd, and L.~Vandenberghe, \emph{Convex optimization}.\hskip 1em
  plus 0.5em minus 0.4em\relax Cambridge university press, 2004.

\bibitem{Losenicky-2018-DissipationFactorsOfSphericalCurrentModes}
V.~Losenicky, L.~Jelinek, M.~Capek, and M.~Gustafsson, ``Dissipation factors of
  spherical current modes on multiple spherical layers,'' \emph{IEEE
  Transactions on Antennas and Propagation}, vol.~66, no.~9, pp. 4948--4952,
  Sept 2018.

\bibitem{Balanis1982}
C.~Balanis, ``Antenna theory: Analysis and design,'' \emph{New York, Harper and
  Row, Publishers, 1982. 805 p.}, 1982.

\bibitem{nec}
T.~C.~A. Molteno, ``{'NEC2++: An NEC-2 compatible Numerical Electromagnetics
  Code'},'' Electronics Technical Reports, Tech. Rep., 2014.

\bibitem{Yaghjian2005}
A.~D. Yaghjian and S.~R. Best, ``Impedance, bandwidth, and {Q} of antennas,''
  \emph{IEEE Transactions on Antennas and Propagation}, vol.~53, no.~4, pp.
  1298--1324, April 2005.

\bibitem{Xu2010}
X.~J. Xu and Y.~E. Wang, ``A direct antenna modulation (dam) transmitter with a
  switched electrically small antenna,'' in \emph{Antenna Technology (iWAT),
  2010 International Workshop on}.\hskip 1em plus 0.5em minus 0.4em\relax IEEE,
  2010, pp. 1--4.

\bibitem{coilcraft:square-air-coil}
\emph{Square Air Core Inductors}, Coilcraft, Oct. 2015, {Document 720-2}.

\bibitem{liu1976nonlinear}
T.~{Liu} and F.~{Tesche}, ``Analysis of antennas and scatterers with nonlinear
  loads,'' \emph{IEEE Transactions on Antennas and Propagation}, vol.~24,
  no.~2, pp. 131--139, March 1976.

\bibitem{landt1983agard}
J.~A. Landt, ``{Effects of nonlinear loads on antennas and scatterers},'' in
  \emph{In AGARD The Performance of Antennas in their Operational Environ. 22 p
  (SEE N84-12367 03-32)}, Sep. 1983.

\bibitem{palmer2019investigation}
A.~Palmer, ``Investigation of a generalized frequency domain method for
  modeling time-varying loads on antennas,'' Ph.D. dissertation, University of
  Oklahoma, 2019.

\end{thebibliography}
\end{document}